\documentclass[aps,prb,twocolumn,superscriptaddress,showpacs]{revtex4}
\usepackage{graphicx}
\usepackage{dcolumn}
\usepackage{color}
\usepackage{mathrsfs}
\usepackage[breaklinks,colorlinks = true,linkcolor = blue,urlcolor=blue,citecolor=blue]{hyperref}
\usepackage{amsmath}
\usepackage{soul}

\usepackage{amssymb}
\usepackage{epstopdf}
\usepackage{subfigure}
\begin{document}
\title{Ultrasonic investigations of spin-ices Dy$_2$Ti$_2$O$_7$ and Ho$_2$Ti$_2$O$_7$ in and out of equilibrium}
\author{S. Erfanifam}
\affiliation{Hochfeld-Magnetlabor Dresden, Helmholtz-Zentrum Dresden-Rossendorf,
D-01314 Dresden, Germany}
\author{S. Zherlitsyn}
\affiliation{Hochfeld-Magnetlabor Dresden, Helmholtz-Zentrum Dresden-Rossendorf,
D-01314 Dresden, Germany}
\author{S. Yasin}
\affiliation{Hochfeld-Magnetlabor Dresden, Helmholtz-Zentrum Dresden-Rossendorf,
D-01314 Dresden, Germany}
\author{Y. Skourski}
\affiliation{Hochfeld-Magnetlabor Dresden, Helmholtz-Zentrum Dresden-Rossendorf,
D-01314 Dresden, Germany}
\author{J. Wosnitza}
\affiliation{Hochfeld-Magnetlabor Dresden, Helmholtz-Zentrum Dresden-Rossendorf,
D-01314 Dresden, Germany}
\author{A.~A.~Zvyagin} 
\affiliation{Max-Planck-Institut f\"ur Physik komplexer Systeme, D-01187 Dresden,
Germany}
\affiliation{B.I. Verkin Institute for Low Temperature Physics and Engineering
of the National Academy of Sciences of Ukraine, Kharkov, 61103, Ukraine}
\author{P. McClarty}
\affiliation{Max-Planck-Institut f\"ur Physik komplexer Systeme, D-01187 Dresden,
Germany}
\author{R. Moessner}
\affiliation{Max-Planck-Institut f\"ur Physik komplexer Systeme, D-01187 Dresden,
Germany}
\author{G. Balakrishnan}
\affiliation{University of Warwick, Department of Physics, Coventry CV4 7AL UK}
\author{O. A. Petrenko}
\affiliation{University of Warwick, Department of Physics, Coventry CV4 7AL UK}
\begin{abstract}
We report ultrasound studies of spin-lattice and single-ion effects in the spin-ice materials Dy$_2$Ti$_2$O$_7$ (DTO)
and Ho$_2$Ti$_2$O$_7$ (HTO) across a broad field range up to 60 T, covering phase transformations, interactions with
low-energy magnetic excitations, as well as single-ion effects. In particular, a sharp dip observed in
the sound attenuation in DTO at the gas-liquid transition of the magnetic monopoles is explained based on an approach involving negative relaxation processes.
Furthermore, quasi-periodic peaks in the acoustic properties of DTO due to non-equilibrium processes are found to be strongly
affected by {\em macroscopic} thermal-coupling conditions: the thermal runaway observed in previous studies in DTO can be suppressed altogether
by immersing the sample in liquid helium. Crystal-electric-field effects having higher energy scale lead to a renormalization
of the sound velocity and sound attenuation at very high magnetic fields. We analyze our observations using an approach based
on an analysis of exchange-striction couplings and single-ion effects.
\end{abstract}
\date{\today}
\maketitle
\section{Introduction}
\label{sec_Introduction}
The rare-earth titanates\cite{Gardner-2010} with magnetic ions on a pyrochlore lattice show a variety of remarkable phenomena particularly (quantum) spin-ice\cite{Ramirez-1999,Bramwell-2001,Castelnovo-2008,Harris-1997,Snyder-2004,Fennell-2009,Morris-2009,Kimura-2013,Chang-2010,Shannon-2012,Powell-2008,Herm-2004} states. For more than a decade, the rare-earth titanates have been intensively studied experimentally and theoretically as model systems of spin ice.

Besides the titanates Dy$_2$Ti$_2$O$_7$ (DTO) and Ho$_2$Ti$_2$O$_7$ (HTO), {$\rm Dy_2Sn_2O_7$} has also been reported to show spin-ice properties.\cite{Kadowaki-2002}
In addition, starting from 2006, artificial spin-ice materials have been developed in lithographically fabricated single-domain ferromagnetic islands.\cite{nisoli-2013}

Back in the 1960s, P.~W.~Anderson considered the direct mapping of the pyrochlore lattice onto water ice.\cite{Anderson-1956}
In the spin-ice state, two spins are directed inward and two spins outward for each tetrahedron which is the building unit of the pyrochlore lattice.\cite{Harris-1997}
The spin-ice ground state has attracted much interest, mainly due to elementary excitations which can be treated as magnetic monopoles for locally fractionalized dipole moments.\cite{Castelnovo-2008,Jaubert-2009,Nakanishi-2011,Giblin-2011}

In addition, low-temperature magnetization measurements of DTO revealed that a magnetic-field-driven first-order phase transition of a gas-liquid type occurs between states with low density and high density of the 3-in 1-out (or 3-out 1-in) configurations.\cite{Sakakibara-2003}
Indeed, these exotic states have been found to display a variety of dynamical features which depend on the specific spin-ice compound.

In recent years, a number of experimental studies have been conducted to explore the non-equilibrium properties of spin ice.
Slobinsky {\em et al.}\cite{Slobinsky-2010} reported monopole avalanches in field-induced magnetization processes in DTO. In references \onlinecite{Kolland-2012,Fan-2013} the dynamics of spin ice was investigated by means of thermal-conductivity measurements. Quantum dynamics of magnetic monopoles in DTO has been recently examined based on results of frequency-dependent magnetic-susceptibility measurements.\cite{Bovo-2013} Tunable nonequilibrium dynamics due to thermal and field quenches has been theoretically studied \cite{2010_castelnovo,2014_mostame} as well as experimentally investigated by means of magnetization measurements.\cite{2014_paulsen} Magnetic-field-driven non-equilibrium processes were also found in our own earlier ultrasound investigations.\cite{Erfanifam-2011}

In order to better understand the physics of spin ice it is vital to perform comparative studies of various spin-ice materials. It is known that  in highly degenerate magnetic systems a lattice distortion can lift the degeneracy leading to an ordered state. This frequently happens in frustrated magnets featured by the pyrochlore lattice and strong spin-lattice interactions.\cite{2002_tchern} Hence, the role of the lattice degrees of freedom, spin-strain interactions, and single-ion magnetoelastic coupling in spin-ice materials have to be throughly investigated. These issues are addressed in this work by performing  a comparative study of  two materials with the spin-ice ground state, Dy$_2$Ti$_2$O$_7$ and Ho$_2$Ti$_2$O$_7$.

Ultrasound is a direct probe for various lattice instabilities and their couplings to spin and orbital degrees of freedom.\cite{2005_luthi} In this paper, measurements of the sound velocity and the sound attenuation are presented for selected acoustic modes propagating in DTO and HTO. Such ultrasound studies are ideally suited\cite{2005_luthi} to better understand the influence of spin-lattice effects. The obtained results provide new insight into the physics relevant in spin-ice materials. The acquired data are analyzed using an exchange-striction approach as well as taking into account crystal-electric-field (CEF) effects. We have performed numerical simulations to investigate the inter-ionic magnetic interactions and spin-phonon coupling.

This paper is organized in the following way: in the next section we provide some details of the sample preparation and experimental setup.
After presenting and describing the experimental data in section \ref{sec_RESULTS}, the corresponding models and numerical simulations are discussed in section \ref{sec_THEORETICAL}.

\section{EXPERIMENT}
\label{sec_EXPERIMENT}
Single crystals were grown by the floating-zone method using an infrared furnace.\cite{Balakrishnan-1998} Dy$_2$Ti$_2$O$_7$ and Ho$_2$Ti$_2$O$_7$ have
a cubic crystal structure with the space group $Fd\overline{3}m$. The samples were cut and polished with two parallel surfaces, perpendicular to the crystallographic directions [111], [112], and [001]. The thickness of the Dy$_2$Ti$_2$O$_7$ samples used in the experiments are 2.57~mm in [111], 3.41~mm in [001], and 1.09~mm in [112] direction, respectively. Ho$_2$Ti$_2$O$_7$ sample is 5.74 mm in [111] and 10.32 mm in [001] direction, respectively. Correspondingly, all samples have a size of a few millimeters in the plane normal to the sound-propagation direction. The alignments were checked for all samples using the x-ray Laue technique. Four acoustic modes have been studied: (i) the  longitudinal $c_{11}$ mode ({\bf k}$\|${\bf u}$\|$[001]), (ii) the longitudinal $c_L$ mode, $c_{L} = (c_{11}+2c_{12}+4c_{44})/3$ ({\bf k}$\|${\bf u}$\|$[111]), (iii) the quasi-longitudinal $c_L^{\prime}$ mode ({\bf k}$\|${\bf u}$\|$[112]), (iv) the transverse $c_T$ mode, $c_T = (c_{11}+c_{44}-c_{12})/3$ ({\bf k}$\|$[111], {\bf u}$\bot${\bf k}). Here, {\bf k} and {\bf u} are the wave vector and polarization of the acoustic wave, respectively. See also the inset of Fig.~\ref{PRB7} where various crystallographic directions mapped on a single tetrahedron of the pyrochlore structure are shown. The elastic constant $c_{ij}$ is related to the sound velocity $c=\rho v^{2}$, where the mass density, $\rho$, is 6.87 g/cm$^{3}$ for Dy$_2$Ti$_2$O$_7$ and 6.94 g/cm$^{3}$ for Ho$_2$Ti$_2$O$_7$, respectively. It is worth noting that we averaged over 10 sound echoes to obtain the time delays which were used for the calculation of the absolute sound velocity and corresponding elastic constants. Pulsed-field magnetizaion was measured by integrating the voltage induced in compensated coils surrounding the sample.  See Ref.~\onlinecite{2011-skou} for detailed set-up description.
\begin{figure}
\begin{center}
\includegraphics[width=0.85\columnwidth]{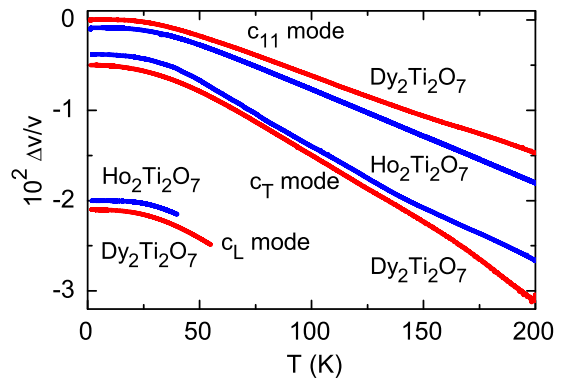}
\end{center}
\caption{(Color online) Temperature dependence of the sound-velocity changes, $\Delta v/v$, in HTO and DTO for  longitudinal, $c_{11}$, (upper two curves),  $c_{L}$ (lower two curves), and transverse, $c_T$, (in the middle) acoustic modes.
The curves are offset along the $y$ axis for clarity.
For DTO, the ultrasound frequencies of 97.7, 96.5, and 171~MHz were used for the $c_{11}$, $c_{L}$, and $c_T$ modes, respectively. For HTO, 50, 49, and 109.5~MHz were used for $c_{11}$, $c_{L}$, and $c_T$, respectively.}
\label{PRB000}
\end{figure}

Calibrated RuO$_{2}$ and Pt100 resistors, thermally coupled with the sample were employed for thermometry. The relative change of the sound velocity, $\Delta v/v$, and the sound attenuation, $\Delta \alpha$, were measured using a phase-sensitive-detection technique.\cite{2005_luthi}
The measurements have been performed under zero-field-cooled (ZFC) conditions. For that, we demagnetized the magnet at $\approx 2.5$~K before cooling to the desired temperature. We used a $^3$He cryostat (0.29 - 7~K) with the samples placed in vacuum but attached to the $^3$He chamber, a dilution refrigerator (0.02 - 2.5~K) with the samples placed in the liquid, as well as a variable temperature insert (VTI, 1.5 - 300~K) with gas or liquid environment for the sample. Static magnetic fields up to 20~T were available by commercial superconducting magnets. The pulsed-field experiments in fields up to 60~T were performed using a gas-flow $^4$He cryostat (1.5 - 300~K). Two pulsed magnets have been employed with the rise time of 33 ms and the whole pulse duration of 150 ms in the ultrasound experiments as well as with 7 ms rise time and 25 ms the whole pulse duration for the magnetization measurements.

\section{RESULTS AND DISCUSSION}
\label{sec_RESULTS}
\subsection{Temperature dependence}
\begin{figure}
\begin{center}
\includegraphics[width=0.85\columnwidth]{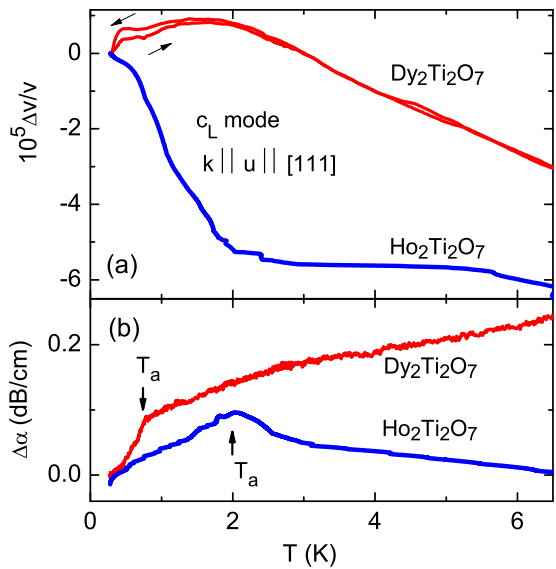}
\end{center}
\caption{(Color online) Temperature dependence of (a) the sound-velocity change, $\Delta v/v$, and (b) the sound attenuation measured at 52~MHz for HTO and 86~MHz for DTO between 0.29~K and 6.5~K. The data at the lowest temperatures were set to zero on the ordinates.
Arrows in (a) indicate temperature-sweep directions.}
\label{PRB003}
\end{figure}

Figures~\ref{PRB000}-\ref{PRB003} show comparative results obtained for DTO (in red) and HTO (in blue) for selected acoustic modes. With decreasing temperature, the sound velocities of DTO and HTO first increase, due to the anharmonicity of the ionic potential\cite{1970_varshni,Nakanishi-2011} and become nearly constant at low temperatures
(Fig.~\ref{PRB000}).
There are some irregular features (change of curvature) for most of the acoustic modes which presumably are caused by crystal-electric-field (CEF) states partly populated at these temperatures.
The absolute values of the sound velocities and elastic constants in DTO and HTO for some acoustic modes at 1.3~K are listed in Table~1.
\begin{table}[ht]
\caption{Absolute values of the sound velocity (elastic constant) in m/s (GPa)  measured at 1.3~K.}
\centering
\begin{tabular}{c c c c c}
\hline\hline
Mode & $v$(DTO) & $v$(HTO) & $c_{ij}$(DTO) & $c_{ij}$(HTO) \\ [0.8ex]
\hline
$c_{11}$ & 7190 & 7142 & 355 & 354 \\
$c_L$ & 6980 & 6705 &  335 & 312\\
$c_T$ & 3670 & 4040 & 92 & 113\\ [2ex]
\hline
\end{tabular}
\label{table:peaks ratio}
\end{table}
The overall features in the acoustic properties at elevated temperatures are very similar for both spin-ice materials. More variations appear below 5 K. Data for the $c_{L}$ mode down to 0.29~K are shown in Fig.~\ref{PRB003}.
In DTO, the sound velocity first grows with increasing temperature up to an
anomaly at temperature, $T_{a} \approx$ 0.5~K, followed by a plateau and a
maximum at approximately 1.7~K.
A small hysteresis is observed here.
Above this temperature, the sound velocity starts to decrease.
In HTO, the sound velocity decreases rapidly from the lowest temperatures
showing a drastic change of the slope near $T_a$ around 2~K. The details of \
the sound-velocity and attenuation change close to and below $T_{a}$ depend
sensitively on the temperature-sweep rate and sample history. Remarkably, both
materials show rather small velocity change just above $T_a$, between
$T_{a}$ and 3$T_{a}$, whereas below $T_{a}$ there is a stiffening of the
$c_{L}$ mode in HTO and a softening of the same mode in DTO.

We note that in a temperature window around $T_a$, both DTO and HTO also
show a well-known, albeit still somewhat enigmatic, onset of dynamical slowing
down, including the appearance of history dependence.\cite{Snyder-2004,Quilliam-2011,Clancy-2009,Bramwell-2001,Jaubert-2011,Krey2012} In DTO, this is
well-established to occur around 0.5 - 0.6~K, whereas in HTO, the situation is
less clear, with a recent study\cite{Krey2012} finding history dependence in
the magnetization setting in at a temperature as small as 0.6~K.
In both compounds, the sound attenuation increases first followed by an
anomaly at the distinguished point (a kink in DTO and a maximum in HTO) and a
smooth change towards higher temperatures (Fig.~\ref{PRB003}b).
In case of HTO, the hysteresis in the vicinity of the distinguished point is
less pronounced (not shown).
Both, the rapid change in the sound velocity and the decrease in the sound
attenuation below the distinguished temperature show that the magneto-elastic interactions are relevant for physics of these spin-ice materials.

\begin{figure}
\begin{center}
\includegraphics[width=0.85\columnwidth]{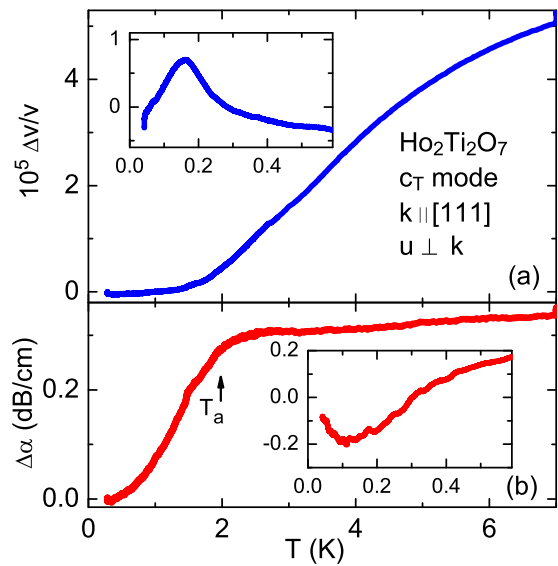}
\end{center}
\caption{(Color online) Temperature dependence of (a) the sound velocity, $\Delta v/v$, and (b) the sound attenuation of the transverse ultrasonic wave propagating along the [111] direction (acoustic $c_T$ mode) in HTO. The ultrasound frequency was 181~MHz. The insets show low-temperature data at 28~MHz obtained using the dilution refrigerator.}
\label{PRB1a}
\end{figure}

Interestingly, the acoustic properties for the transverse mode propagating along the [111] direction behave quite different from the $c_L$ mode. Data for HTO are shown in Fig.~\ref{PRB1a}.
Remarkably, the sound velocity decreases  with temperature reduction. At 2~K, a clear change in the slope both in the sound velocity and in the sound attenuation occurs.
Apparently, due to a stronger spin-lattice interaction and coupling to the spin degrees of freedom, the $c_{T}$ mode begins to soften at temperatures much higher than the $c_L$ mode (Fig.~\ref{PRB003}).
The insets of Fig.~\ref{PRB1a} show the sound velocity and attenuation measured in the dilution refrigerator below 0.6~K.
Well pronounced anomalies in both acoustic characteristics appear at about 0.15~K.
The origin for this is not clear at present.
Note, that possible ordered phases at very low temperatures in spin-ice materials have been discussed previously.
\cite{Melko-2001,Gaulin2014} However, definite conclusion cannot be drawn from the ultrasound data. Anyhow, from our results it seems reasonable to assume that there is no structural transformation in HTO at the lowest temperatures, since no abrupt, temperature-localized features expected for the acoustic modes in case of structural transformation have been detected.
\subsection{Magnetic-field dependence}
In a previous work,\cite{Erfanifam-2011} we reported some striking anomalies in the sound velocity and the sound attenuation in DTO with sweeping the magnetic field. We ascribed that to non-stationary processes in the spin-ice state.
The appearance of these processes strongly depends on the adiabatic conditions during the experiment.
To explore the influence of the heat exchange between sample and environment, we performed experiments for two different sample-bath coupling conditions.
In the first experiment, the sample was located in the $^3$He-$^4$He liquid of a dilution refrigerator (strong thermal coupling) and in the second experiment, the sample was attached to a $^{3}$He chamber via a thermal link. The latter case leads to a weaker coupling to the bath, still providing a reliable temperature control of the sample.
We further took into account the demagnetization factors of our samples to trace how they affect the obtained results. Note, that in all cases, a gas-liquid-type transition  between the high- and low-density phases of magnetic monopoles without any change in the symmetry has been observed in the magnetic field applied along the [111] direction.\cite{Aoki-2004,2009_kadowaki,2006_tabata,Matsuhira-2007,Erfanifam-2011}
\begin{figure}
\begin{center}
\includegraphics[width=0.85\columnwidth]{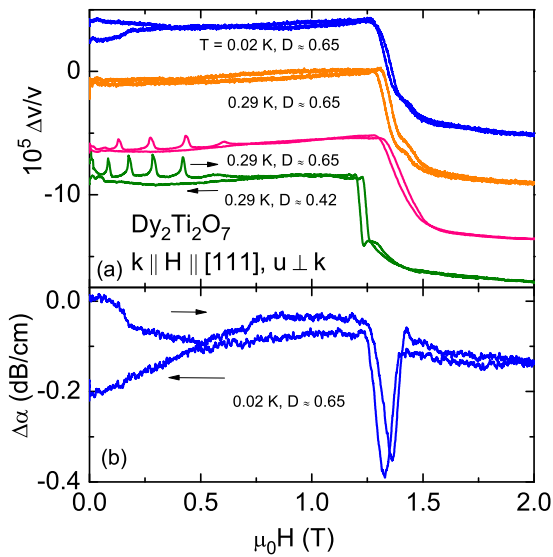}
\end{center}
\caption{(Color online) Field dependence of (a) the sound velocity, $\Delta v/v$, and (b) the sound attenuation of the $c_T$ mode. The upper two curves in (a) were obtained in the dilution refrigerator, whereas the lower two curves were measured in the $^{3}$He cryostat (see text for details).
Up and down sweeps are indicated by arrows.
The ultrasound frequencies were 108, 108, 183 and 170~MHz from the upper to lower curve, respectively. The field sweep-rate was 0.015~T/min except for the sample with $D$ = 0.65 at 0.29~K, which was 0.03~T/min. The curves are arbitrarily shifted along the $y$ axis for clarity.}
\label{PRB}
\end{figure}

In Fig.~\ref{PRB}, the sound velocity and sound attenuation of the acoustic $c_T$ mode in two DTO samples with different demagnetization factors and for the described bath-coupling conditions are shown as a function of magnetic field.
The data were collected at 20~mK (strong thermal coupling) and 0.29~K (both strong and weak thermal coupling).
In addition, at 0.29~K, data for two samples with two different demagnetization factors, $D = 0.42$ (sample 1, also used in Ref. \onlinecite{Erfanifam-2011}) and 0.65 (sample~2) are shown.
The second sample was prepared from the first one by cutting in half the crystal parallel to the (111) plane.
The corresponding demagnetization factors are estimated by approximating the sample to the closest rectangular shape and for the magnetic field applied along the [111] axis.
\begin{figure}
\begin{center}
\includegraphics[width=0.85\columnwidth]{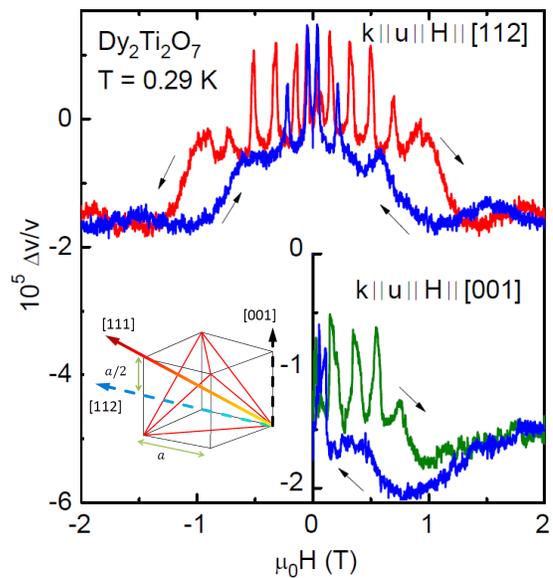}
\end{center}
\caption{(Color online) Field dependence of the sound velocity,~$\Delta v/v$, for different field and sound propagation directions at 0.29~K under weak thermal coupling conditions. The upper curve shows $\Delta v/v$ for a quasi-longitudinal ultrasonic wave propagating along [112] ($c^\prime_L$ mode). The magnetic field first was swept to 2~T, then to -2~T and back to 0. The low curve shows $\Delta v/v$ for the $c_{11}$ mode propagating along [001].
The sound frequencies for  $c_{11}$ and $c^\prime_L$ mode were 64.5 and 86~MHz, respectively. Inset: Various crystallographic directions mapped on a single tetrahedron of the pyrochlore structure are shown.}
\label{PRB7}
\end{figure}

For the weaker thermal coupling [lower two curves in Fig.~\ref{PRB}(a)], a number of quasi-periodic peaks in the sound velocity appear. As discussed earlier, \cite{Erfanifam-2011} these peaks signal non-equilibrium processes in the spin-ice state. For the different demagnetization factors, the peaks are somewhat shifted in field and reduced in amplitude (mainly due to the different masses and heat capacities), but the overall picture is unchanged. The main effect of the increased demagnetization factor is the expected slight increase of the gas-liquid-type transition in fields from about 1.25~T for $D = 0.42$ to approximately 1.35~T for $D = 0.65$.

The non-equilibrium peaks disappear when sample 2 is strongly coupled to the thermal bath [upper two curves in Fig.~\ref{PRB}(a)]. Obviously, here the heat release caused by the field-induced monopole avalanches is thermalized much faster than our sound-velocity measuring time. This establishes that heat loss to the environment is a crucial bottleneck responsible for the thermal runaway, although in this sense is not intrinsic to the spin-ice magnetic system. Note,  that our ultrasound measurements provide a very direct probe of the bulk sample conditions, unlike a thermometer attached to the sample surface.

At the lowest temperature (0.02~K), we observe a pronounced hysteresis below 0.3~T, that is absent at 0.29~K [Fig.~\ref{PRB}(a)]. A strong hysteresis also appears in the sound attenuation at 0.02~ K [Fig. \ref{PRB}(b)]. In addition, a sharp and quite unique dip occurs at the gas-liquid transition at 1.35~T. This may be considered as a kind of sound amplification and is discussed in Section~IV.
\begin{figure}
\begin{center}
\includegraphics[width=0.85\columnwidth]{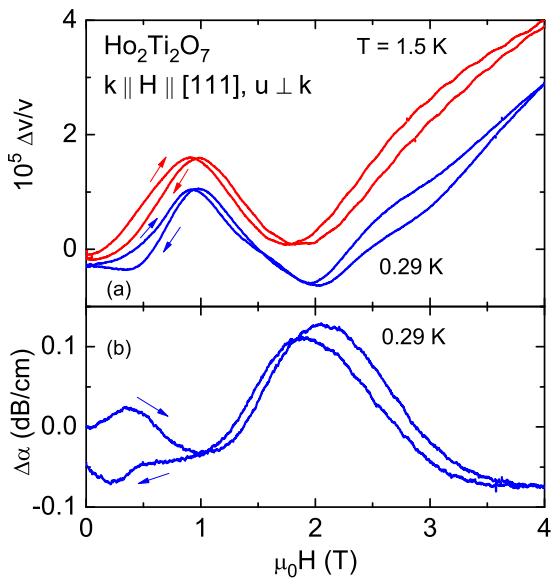}
\end{center}
\caption{(Color online) Field dependence of (a) the sound velocity, $\Delta v/v$, and (b) sound attenuation, $\Delta \alpha$, of the transverse $c_T$ mode propagating along the [111] direction for 0.29 and 1.5~K.
The ultrasound frequency was 181~MHz. Arrows indicate field-sweep directions.}
\label{PRB8}
\end{figure}

In order to explore the nature of the field-induced phase transition and the non-equilibrium features appearing in the spin-ice state further, we have measured other acoustic modes in DTO.
Figure~\ref{PRB7} shows the field-dependent sound velocity at 0.29~K of the $c_{11}$ and $c^\prime_L$ modes for the magnetic field applied along [001] and [112], respectively. For the latter case, magnetic field was first swept up to 2~T, then reversed to -2~T and back to 0. The experimental curves are very symmetric with respect to $H = 0$.
For both acoustic modes, large hystereses are found, although the sharp anomalies at the gas-liquid-type transition observed for $H \| [111]$ (Fig.~\ref{PRB}) are missing. The non-equilibrium peaks, however, are nicely observed also for these modes and field directions for the weak thermal-coupling condition ($^{3}$He cryostat). Indeed, the same number of peaks appear in the up sweeps for all field directions.

For comparison, we investigated the field-dependent sound velocity and attenuation of the $c_{T}$ mode in HTO as well (Fig.~\ref{PRB8}). For 0.29 and 1.5~K, the sound velocity exhibits a maximum at 1~T and a minimum at higher fields. This minimum shifts from about 2~T at 0.29~K to 1.8~T at 1.5~K.

Concomitantly, the sound attenuation shows a noticeable peak at about 2~T at 0.29~K associated with the gas-liquid-type transition.
The corresponding critical fields are in good agreement with the transition obtained by Fennell {\em et al.}\cite{Fennell-2009} from neutron-scattering and magnetization measurements performed at 0.51~K.
Sharp slope changes in the magnetization are also observed at approximately the same magnetic field.\cite{Petrenko-2003,Petrenko-2011}

In HTO, however, we do not observe any non-equilibrium peaks due to monopole avalanches. If present, relaxation to equilibrium is too fast to be detected in our experiment. Taking into account the field-sweep rate we can assume that the relaxation to equilibrium in HTO is shorter than 1 s being much faster than estimated 1 min in DTO.\cite{Erfanifam-2011} This could also be the reason for the significant differences seen for the $c_{T}$ mode at the gas-liquid-type transition in DTO and HTO. This question demands further investigation.
\begin{figure}
\begin{center}
\includegraphics[width=0.85\columnwidth]{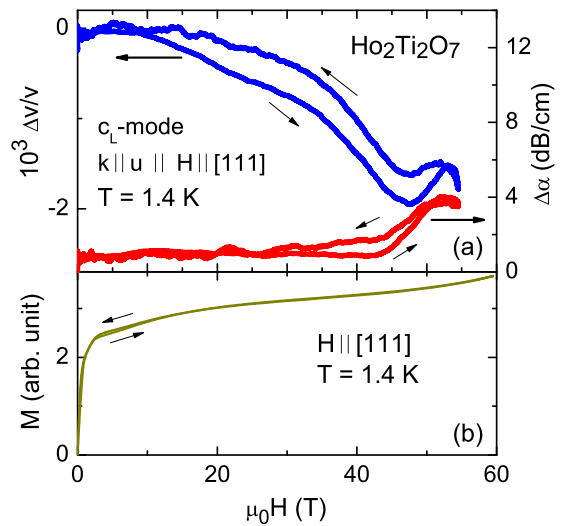}
\end{center}
\caption{(Color online)~Field dependence of (a) the sound velocity, $\Delta v/v$, and sound attenuation, $\Delta \alpha$, for the $c_L$ mode and (b) the magnetization in HTO measured at 1.4~K.
The ultrasound frequency was 70.9~MHz. Thin arrows indicate field-sweep directions.}
\label{pulsehto}
\end{figure}
\subsection{Crystal-electric-field effects}
Here, we consider the role of the CEF and single-ion-strain interactions in the spin-ice materials. The CEF contributes to the single-ion anysotropy having much larger energy scale in comparison to the exchange and dipole-dipole interactions in DTO and HTO. Hence, high magnetic fields are required for these investigations. We performed ultrasound and magnetization experiments in pulsed magnetic fields up to 60~T.
Figure~\ref{pulsehto} shows the field dependence of the sound velocity and sound attenuation ($c_{L}$ mode) as well as of the magnetization in HTO at 1.4~K.
The sound velocity decreases with increasing  magnetic field and pronounced anomalies appear in the sound velocity and sound attenuation around 50~T. An upturn in the magnetization is observed in the same field range. These anomalies can be attributed to CEF energy-level crossings (see section~\ref{ss:singleiontheory}).
For DTO, the magnetization exhibits a plateau around 2~T in the up sweep accompanied by a pronounced hysteresis (not shown).
The high-field data show that the magnetization in these compounds does not saturate even at 60~T (see the discussion in section~\ref{sec_THEORETICAL}).
\begin{figure}
\begin{center}
\includegraphics[width=0.85\columnwidth]{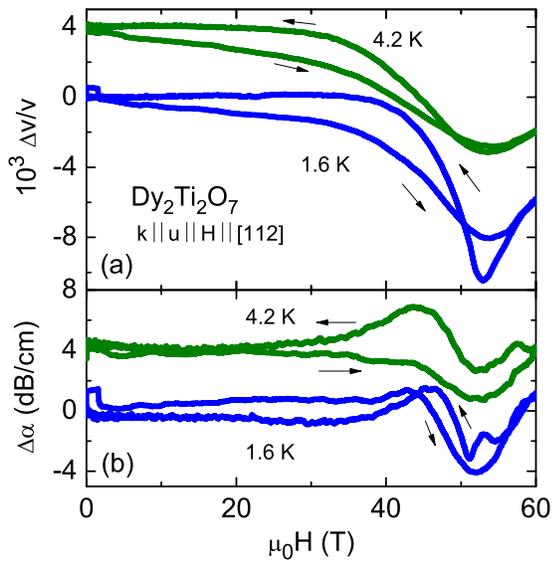}
\end{center}
\caption{(Color online)~Field dependence of (a) the sound velocity, $\Delta v/v$, and (b) sound attenuation, $\Delta \alpha$, of the $c^\prime_L$ mode in DTO at 1.6 and 4.2~K. The ultrasound frequency was 53.4~MHz. Arrows indicate field-sweep directions. The data for different temperatures are shifted along the $y$ axis for clarity.}
\label{dto112puls}
\end{figure}

In Fig.~\ref{dto112puls}, the field dependence of the sound velocity and sound attenuation of the $c^\prime_L$ mode in DTO is shown.
Strong field-dependent changes appear above 35~T.
Both, the sound velocity and attenuation, show strong hystereses and pronounced minima at about 53~T.
These minima are sharper at the lower temperature of 1.6~K compared to 4.2~K.
They are discussed in section~\ref{sec_THEORETICAL} by a CEF model. The observed hysteresis in both materials is apparenly attributed to the magnetocaloric effect demanding additional study.
\section{THEORETICAL ANALYSIS}
\label{sec_THEORETICAL}

This section is devoted to a theoretical analysis accompanying the measurements discussed in the previous sections. Our analysis concerns two regimes, firstly the low-field regime, in which cooperative physics dominates; and secondly, the high-field regime, where the equivalent energies are so large that they probe single-ion physics in the form of crystal-field levels above the lowest doublet. 

We first provide a very brief overview to the basic framework for analysing ultrasound data to probe spin physics in Sec.~\ref{ss:spinphonontheory}
-- for a more exhaustive account of technical details, we refer the reader to Ref.~\onlinecite{2005_luthi}. This requires the input of thermodynamic quantities for the interacting spin system, which we extract from a Bethe-Peierls approximation (Sec.~\ref{ss:bethepeierls}), to yield sound velocity renormalization and attenuation in the low-field regime (Sec.~\ref{ss:deltau_theory}). Sec.~\ref{ss:singleiontheory} provides the requisite analysis of the field dependence of the crystal level scheme, in order to address some qualitative features of the high-field regime.

\subsection{Effect of spin-phonon coupling on the exchange integrals in spin ice}
\label{ss:spinphonontheory}

The renormalization of phonon velocities and sound attenuation in magnetic media for stationary situation is caused mostly by two reasons. \cite{2005_luthi} First, sound waves change ligand positions, and, therefore, the crystalline electric field of ligands is affected. Due to strong spin-orbit interaction, the latter produces changes of the single-ion magnetic anisotropy of rare-earth ions. This should also change the effective $g$-factors of rare-earth ions. Thus, due to magneto-elastic interactions, sound waves can change the directions of the magnetic anisotropy slightly, and vice versa, the magnetic anisotropy modifies the sound-wave parameters, such as the sound velocity and attenuation. This interaction between the sound wave and the magnetic properties exists at all temperatures
lower than the characteristic energy of the single-ion magnetic anisotropy.

On the other hand, sound waves change the magnetic ion positions themselves and/or the positions of nonmagnetic ions involved in the superexchange between magnetic ones. In this case, sound waves renormalize the effective exchange (as well as dipolar) interactions between magnetic ions. The manifestation of this effect is more pronounced compared to the influence of the sound on the single-ion magnetic properties, because inter-ionic magnetic and exchange interactions predominantly determine magnetic phase transitions. Here the relative renormalization of the sound velocity $\Delta v/v$ and the sound attenuation $\Delta \alpha$ are related to the changes of the exchange integrals due to effect of sound waves on distances between the moments involved into the exchanges. 
These renormalizations are proportional to spin correlation functions; the latter can be approximated as combinations of the components of the magnetization and magnetic susceptibility of the magnetic subsystem. 

In DTO and HTO, the single-ion anisotropy is of the order of 300~K. This is why we expect that the sound-mode changes at low temperatures are mostly related to the renormalization of inter-ionic magnetic interactions due to the spin-phonon coupling. To describe the features of the sound in DTO and HTO we adapt the approach used in Refs.~\onlinecite{Chia,Sych}. 

The relative renormalization of the longitudinal sound velocity can be written as (in this subsection we use the notations in which $g_{\rm J}\mu_{\rm B} =k_{\rm B} =\hbar =1$)
\begin{equation}
{\Delta v\over v} = - {(A_1 +A_2)\over (N\omega_{\bf k})^2} \ ,
\label{otn}
\end{equation}
where
\begin{eqnarray}
&&A_1 = 2 |G_0({\bf k})|^2 \langle S_0 \rangle^2 \chi_0
\nonumber \\
&&+ T\sum_{\bf q} \sum_{\alpha =x,y,z} |G_{\bf q}^{\alpha} ({\bf k})|^2
(\chi_{\bf q}^{\alpha})^2 \ ,
\nonumber \\
&&A_2 = H_0({\bf k}) \langle S_0 \rangle^2 +{T\over
2}\sum_{\bf q} \sum_{\alpha =x,y,z} H_{\bf q}^{\alpha} ({\bf k})
\chi_{\bf q}^{\alpha} \ .
\label{otn1}
\end{eqnarray}
Here $N$ is the number of spins in the system, $\omega_{\bf k} =v k$ is the low-$k$ dispersion relation for the phonon with the sound velocity $v$ in the absence of spin-phonon interactions, $\langle S_0 \rangle$ is the average spin moment (related to the magnetization) along the direction of the magnetic field, $H$, $\chi_{\bf q}^{x,y,z}$ are non-uniform magnetic susceptibilities, and the subscript $0$ corresponds to $q=0$. 
The renormalization is proportional to the spin-phonon coupling constants $G_{\bf q}^{\alpha}$ and $H_{\bf q}^{\alpha}$ (which have to be determined independently). The latter are proportional to gradients of exchange integrals. We use these 
constants as fitting parameters in the following. 

Following Refs.~\onlinecite{Chia,Sych} we can calculate the attenuation coefficient
\begin{eqnarray}
&&\Delta \alpha (\equiv \Delta \alpha_k) ={1\over Nv} \bigg[2
|G_0({\bf k})|^2 \langle S_0 \rangle^2 \chi_0
{\gamma_0^z\over (\gamma_{0}^z)^2 + \omega_{\bf k}^2}
\nonumber \\
&&+ T \sum_{\bf q} \sum_{\alpha =x,y,z} |G_{\bf q}^{\alpha} ({\bf
k})|^2 (\chi_{\bf q}^{\alpha})^2{2\gamma_{\bf q}^{\alpha}\over
(2\gamma_{\bf q}^{\alpha})^2 + \omega_{\bf k}^2} \bigg] \ , \
\label{alpha}
\end{eqnarray}
where $\gamma_{\bf q}^{\alpha}$ are the relaxation rates, which can be approximated by $\gamma_{\bf q}^{\alpha} =B/T\chi_{\bf q}^{\alpha}$, where $B$ is a material-dependent constant (see Ref.~\onlinecite{TM}). 
\subsection{Calculation of the magnetization and magnetic susceptibility of spin ices}
\label{ss:bethepeierls}
According to the above approach, the renormalization of the sound velocities and attenuation due to the magneto-elastic interaction, is related to the behavior of the magnetization (total spin moment) and the magnetic susceptibility, affected by the effective exchange interactions between the rare-earth magnetic ions in DTO and HTO. We can calculate the temperature and magnetic field dependences of the total spin moment and magnetic susceptibility of DTO and HTO with a help of a recently developed approach, \cite{Tim,Jaub-08} in which a
Bethe-Peierls analysis on a Bethe lattice  is used, in analogy to our recent work on the spinels.\cite{bhat-2011} The Bethe-Peierls approach allows considerable progress in obtaining analytical expressions for the magnetic susceptibility and magnetization of studied spin ices.

In that approach the free energy of the spin-ice system can be written as
\begin{equation}
F= {T\over 4} \sum_{\alpha=0}^{3} \ln [2\cosh(2f_{\alpha} -h_{\alpha})] -{T\over 2} \ln 2Z({\bf f}) \ ,
\label{F}
\end{equation}
where
\begin{equation}
Z({\bf f}) = \sum_{n=0}^2 Z_n({\bf f})e^{-2n^2J/T}.
\label{Z}
\end{equation}

The index $\alpha$ denotes four directions for the easy-axis magnetic anisotropy (considered here to be much larger than the effective interactions between spins) in each tetrahedron in the spin-ice system (with the unit vectors directed along the easy axes
${\bf e}_{0,1} = ({\bf x} \pm {\bf y} \pm {\bf z})/\sqrt{3}$, ${\bf e}_{2,3} = (-{\bf x} \pm {\bf y} \mp {\bf z})/\sqrt{3}$, where
${\bf x}$, ${\bf y}$, and ${\bf z}$ are the unit vectors along the co-ordinate axes. $h_{\alpha} \equiv ({\bf H}{\bf e}_{\alpha} )/T$ are
the projections of the external magnetic field $H$ normalized by the temperature, and $f_{\alpha}$ are the projections of the effective magnetic field from other spins, which acts on the spins in the considered tetrahedron. In Eq.~(\ref{Z}) $J$ denotes the effective exchange interaction between spins in each tetrahedron, and
\begin{eqnarray}
&&Z_0({\bf f}) = \cosh (f_0 +f_1-f_2-f_3)
\nonumber \\
&&+2\cosh(f_0-f_1)\cosh(f_2-f_3) \ ,
\nonumber \\
&&Z_1({\bf f}) = \sum_{\alpha=0}^3 \cosh\left( \sum_{\beta=0}^3 f_{\beta}
-2f_{\alpha} \right) \ , \nonumber \\
&&Z_2 = \cosh \left( \sum_{\alpha=0}^3 f_{\alpha} \right) \ ,
\label{Zn}
\end{eqnarray}
are related to the three possible spin configuration in each tetrahedron: two spins directed inside the tetrahedron and two spins directed outside (``two in and two out''); ``three in and one out'' (or vice versa, ``one in and three out''); and ``four in'' (or ``four out''), so that for larger $J$ the most favorable configuration in the absence of the external field is ``two in and two out''. It turns out that the sign of $J$ in the approach is taken such that ``two in and two out'' configuration has the lowest energy. The value of the effective exchange constant, $J$, will be chosen to satisfy the experimental data in DTO and HTO. The projections of the effective field,
$f_{\alpha}$, satisfy the following set of equations
\begin{equation}
\tanh (2f_{\alpha} -h_{\alpha}) = {\partial \ln Z({\bf f}) \over
\partial f_{\alpha}}  \ .
\label{mean}
\end{equation}
The spin moment in this approximation can be written as
\begin{equation}
\langle S_0 \rangle = {1\over 4} \sum_{\alpha=0}^3 {\bf e}_{\alpha}
\tanh (2f_{\alpha} -h_{\alpha}) \ ,
\label{S0}
\end{equation}
and the magnetic susceptibility is
\begin{equation}
\chi_0 = {\partial \langle S_0 \rangle\over \partial H} \ .
\label{susc}
\end{equation}
Obviously, in this approximation we can use only homogeneous contributions from the spin moment and susceptibility to the renormalization of the sound velocity and attenuation.
\subsection{Results for the sound velocity and attenuation}
\label{ss:deltau_theory}

Let us consider the field direction, ${\bf H} = H{\bf e}_0$ 
(i.e. $[111]$, other directions of the external field can be treated in a similar way), 
so that $h_0= H/T$, $h_{1,2,3} = -H/3T$. The results depend \cite{Tim} on the thermal history (``field cooling'' and ``zero-field cooling''). To find the temperature dependence of the sound velocity and sound attenuation in zero magnetic field (i.e., zero-field cooling), we can take into account that under such condition $\langle S_0 \rangle =0$ and $\chi \equiv \chi_0 = 2/3T$, i.e., both the renormalization of the sound velocity and attenuation have to be inversely proportional to the temperature.
\begin{figure}
\begin{center}
\includegraphics[width=0.9\columnwidth]{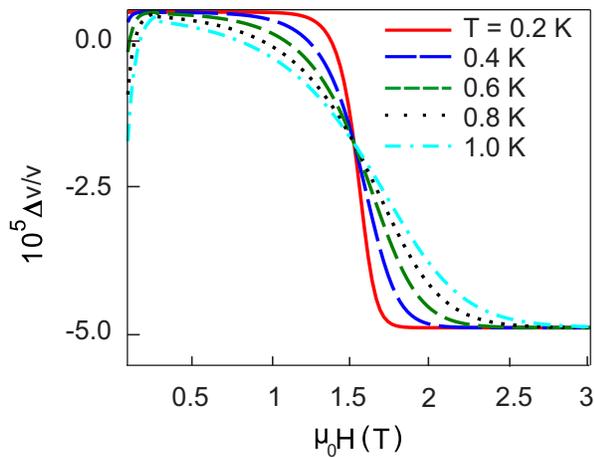}
\end{center}
\caption{(Color online) Calculated magnetic-field dependence of the relative change of the sound velocity, $\Delta v/v$, at different temperatures for the field applied along the $[111]$ direction.}
\label{v111H}
\end{figure}
\begin{figure}
\begin{center}
\includegraphics[width=0.9\columnwidth]{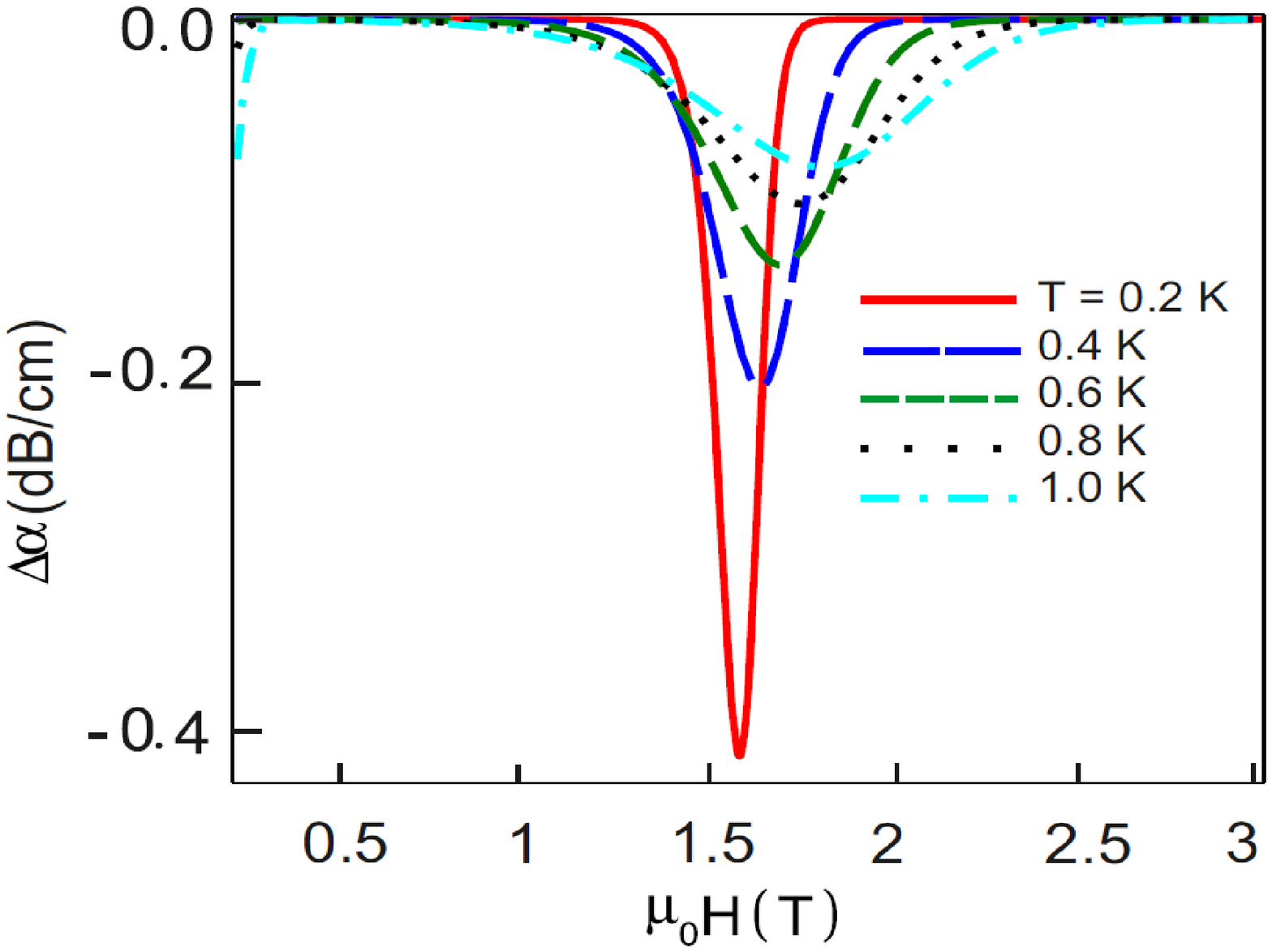}
\end{center}
\caption{(Color online) Calculated magnetic-field dependence of the sound attenuation, $\Delta \alpha$, at different temperatures for the field applied along the $[111]$ direction.}
\label{pogl111H}
\end{figure}

A qualitative explanation of the ``negative relaxation'' (i.e. $B <0$ in the expression for the relaxation rate) can be the following. For spin-ice systems in the absence of an external field, the ground state is highly degenerate. Under the influence of the sound the degeneracy can be locally lifted. It is possible, then, that the relaxation is not to the configuration with the lowest energy, which has to be positive, but to some dynamical steady-state configuration of spins with energy higher than the lowest energy of spins in the absence of sound (cf. Ref.~\onlinecite{Zv}, where the relaxation of parametrically excited magnons in the spin system was to the dynamical
steady-state configuration of spins, which had higher energy in the absence of parametric pumping), which can yield effective
``negative relaxation''.

The ``field-cooled'' case, instead, implies for the field directed along $[111]$,  $\chi=0$ and $\langle S_0 \rangle =1/3$, i.e. an almost temperature-independent renormalization of the sound velocity and zero sound attenuation at low temperatures for $H \to 0$.

Figures~\ref{v111H} and \ref{pogl111H} show the calculated magnetic-field dependences of the sound velocity and attenuation at several temperatures. The sound velocity shows a step-like feature at a critical magnetic field, while the attenuation shows a deep minimum, in agreement with our experiment (Fig.~\ref{PRB}).

At this value of the external magnetic field, directed along $[111]$, the magnetization jumps from $1/3$ to $1/2$ of the saturation value. As the temperature increases, it broadens these features, slightly shifting their positions to higher fields. The features near $H =0$ are related to the step-like field-induced magnetization, observed in DTO and HTO. This, in fact, is the the manifestation of the transition between the spin-ice and Kagome-ice phases for the external magnetic field directed along $[111]$ (see, e.g. Ref.~\onlinecite{Zv-2013}). In the absence of field, all states are degenerate (with the lowest-energy configuration ``two in - two out''), which yields the large residual Pauling entropy ${\cal S}=(1/2)\ln(3/2)$ per spin. On the other hand, the small nonzero magnetic field fixes the ``out'' direction of one spin, while the three other spins in the tetrahedron are free to choose, which of them is directed ``out'' to fulfill the spin-ice rule. In this case, the
entropy of the system becomes smaller than the Pauling value, ${\cal S} = (1/4)\ln(4/3)$ in fact, the exact value is ${\cal S}= $0.08077.\cite{Moes-03}

To summarize, our simplified theory explains well the magnetic field behavior of the sound velocity and attenuation. The temperature behavior in the external magnetic field (the field-cooled case), given by the theory, also qualitatively agrees with the experimental results. However, the zero field-cooled case, i.e. the temperature behavior of the sound characteristics in the absence of the external field does not agree well with the data of our experiments. Such a disagreement is, probably, related to the fact that Bethe-Peierls approximation for spin ices works less satisfactory for absent fields, where the highest level of degeneracy in spin ices is expected.

\subsection{Single-ion magnetoelastic coupling}
\label{ss:singleiontheory}
The purpose of this section is to provide an explanation for the pronounced changes in the sound velocity at high fields applied along
the $[111]$ (Fig.~\ref{pulsehto}) and [112] (Fig.~\ref{dto112puls}) directions. We do not analyze here the data shown in Fig.~\ref{PRB000}  because of a big anharmonic background in the temperature dependence of the sound velocity, which cannot be accurately subtracted from the data. The fields attained by the experiments are on the scale of
$\mu_{\rm B}\langle J\rangle H/k_{\rm B}$ of order of 100~K which is much larger than the scale of the interactions (which are of the order of
$1$ K) but comparable to the CEF splittings. Therefore, in contrast to the previous section, we ignore the exchange couplings and
consider instead the coupling between the lattice degrees of freedom and the single-ion anisotropy mediated by deformations of the
crystal-field environment surrounding the magnetic ions.
The model we study is
\[   H = H_{\rm el} + H_{\rm CF} + H_{\rm Z} + H_{\rm mag-el}.   \]
The first term, $H_{\rm el}$, is the energy cost of deforming the crystal. The second part $H_{\rm CF}$, is the single-ion crystal-field
Hamiltonian on each site with Zeeman term $H_{\rm Z}$. The magneto-elastic interaction is $H_{\rm mag-el}$.

Since the materials are cubic there are three independent elastic moduli by symmetry so,
\begin{eqnarray}
&&H_{\rm el} = \frac{1}{2}c_{11} \left(\epsilon_{xx}^{2} + \epsilon_{yy}^{2} + \epsilon_{zz}^{2}  \right) \nonumber \\
&&+ c_{12} \left(\epsilon_{xx}\epsilon_{yy} + \epsilon_{xx}\epsilon_{zz} + \epsilon_{zz}\epsilon_{yy}  \right) \nonumber \\
&&+ 2c_{44} \left( \epsilon_{xy}^{2} +  \epsilon_{xz}^{2} +  \epsilon_{yz}^{2}   \right) \ ,
\end{eqnarray}
where the strains $\epsilon_{\alpha\beta}$ are dimensionless and the elastic moduli $c_{\alpha\beta}$ expressed in GPa.
The crystal field and Zeeman terms are
\begin{eqnarray}
&&H_{\rm CF} = \sum_{i} B^{2}_{0}O^{2}_{0}(i) + B^{4}_{0}O^{4}_{0}(i)
\nonumber \\
&&+  B^{4}_{3}O^{4}_{3}(i) + B^{6}_{0}O^{6}_{0}(i)
+ B^{6}_{3}O^{6}_{3}(i) + B^{6}_{6}O^{6}_{6}(i) \ , \nonumber \\
&&H_{\rm Z} = - g_{\rm J} \mu_{\rm B} \sum_{i} \mathbf{J}_{i}\cdot \mathbf{H} \ . \label{eqn:HCF} 
 \end{eqnarray}
Here, $O^{l}_{m}(i)$ are Stevens operators on site $i$. The crystal-field Hamiltonian is written in local coordinate frames - one for each
of the four sublattices in the tetrahedral basis - in which the local $z$ axes are aligned along the respective [111] directions, see
above. The Stevens operators are restricted to $l$ = 2,4 and 6 because the magnetic ions have orbital angular momentum $L=3$. The further
restriction to the six operators given above comes about because the crystal-field environment is symmetric under the operations of point
goup $D_{3d}$. In the following, we take parameters $B^{l}_{m}$ for HTO ($J_{\rm Ho}=8$) obtained by fitting the spectrum of
Eq.~(\ref{eqn:HCF}) to inelastic neutron data.\cite{Rosenkranz} For DTO ($J_{\rm Dy}=15/2$), we convert the aforementioned parameters
assuming a point-charge approximation for the crystal field
\begin{equation}  B^{l}_{m}({\rm DTO}) = B^{l}_{m}({\rm HTO}) \frac{\Theta_{l,{\rm Dy}} \langle r^{l} \rangle_{\rm Dy} }{\Theta_{l,{\rm Ho}}
\langle r^{l} \rangle_{\rm Ho} } \ ,  \label{eqn:convert} \end{equation}
where $\langle r^{l} \rangle$ are $4f$ radial expectation values and the $\Theta_l$ are Stevens factors which have been tabulated. In
$H_{\rm Z}$, $g_{\rm J}$ is the Land\'{e} factor appropriate to the magnetic ion under consideration ($g_{\rm J, Ho}=5/4$,
$g_{\rm J, Dy}=4/3$).

The magneto-elastic coupling results from the crystal-field Hamiltonian by expanding in small displacements about the equilibrium ionic
positions. It is convenient to separate these displacements into macroscopic strain, macroscopic rotations, and microscopic strains coming
from displacements of the sublattices. In the following we (i) consider the material to be completely homogeneous and (ii) include only
the effect of macroscopic strains from which the sound velocities are to be computed. Even with these simplifications, the magneto-elastic
couplings are
\begin{eqnarray}
&&H_{\rm mag-el} = \sum_{i} \sum_{l,m} \left[ g^{l}_{m}
\right]_{\alpha\beta}\bar{\epsilon}_{\alpha\beta} O^{l}_{m}
\nonumber \\
&&= \sum_{i,\Gamma} \sum_{l,m} \left[ g^{l}_{m}  \right]_{\nu}(\Gamma) \bar{\epsilon}_{\nu}(\Gamma) O^{l}_{m} \ ,
\end{eqnarray}
where $\Gamma$ denotes an irreducible representation (irrepn) of the group of local crystal field symmetry, $D_{3d}$, and $\nu$ runs over
the basis states of the irrepn. The allowed strains can be grouped according to how they transform with respect to the point group
$D_{3d}$ of the local crystal field environment: $A^1_{1g}$ (1D), $A^2_{1g}$ (1D), $E^1_g$ (2D), and $E^2_g$ (2D).

We now compute the change in the longitudinal sound velocity in a magnetic field applied along the $[112]$ direction for DTO. We set the
temperature to $4.2$~K. Whereas in zero field all four sublattices of the pyrochlore lattice are equivalent, the $[112]$ magnetic field
results in three inequivalent directions with the moments on one of the sublattices (in our convention, sublattice $3$) perpendicular to
the field.
We have also examined spectra for sets of parameters computed from first principles using the point-charge approximation. The generic
features of the low-lying crystal field levels are as follows. The degeneracy of the lowest Ising doublet is split at the smallest fields
but only weakly on sublattice $3$ as a result of admixing with excited levels. There are level crossings of the first-excited levels above
the low-lying doublet at the fields probed by experiment and a ground state level crossing on sublattice $3$ between $80$ T and $90$~T
which is associated with a gradual upturn in the magnetization.

We have studied the sound velocity as a function of field for many choices of magnetoelastic couplings. Typically, we observe two
principal features in $\Delta v/v$. The first is a peak or trough which is associated with the thermal depopulation of the first excited
level. We make this correspondence based on the fact that the temperature is set to a value between $1.6$ and $4.2$~K and, at these low
temperatures, the feature is found to shift to lower fields at lower temperatures. In contrast, the experimental data reveals an anomaly
with a minimum at about $52$ T at both $1.6$ and $4.2$~K (Fig. \ref{dto112puls}). The second main feature in our calculation of
$\Delta v/v$ is a large anomaly at the ground-state level crossing that occurs at about $80$~T in the spectrum computed from the
parameters obtained (i) from the point-charge approximation, (ii) by converting parameters from those obtained in
Ref.~\onlinecite{Rosenkranz}, and (iii) the parameters reported in Ref.~\onlinecite{Bertin}. However, these sets of crystal field
parameters are not taken directly from experimental measurements. One might consider the sound-velocity data reported here to indicate a
ground state level crossing on sublattice~1 at around $60$~T. If this is the case then the available crystal field parameters either
overestimate the crystal field gap or underestimate the moment of the first excited levels.

For a field aligned along the [111] direction in HTO the longitudinal sound velocity decreases showing a minimum at about 47~T
(Fig. \ref{pulsehto}). This field direction has the characteristic of creating two inequivalent magnetic sites - one with the field
aligned along the direction of the zero-field anisotropy (sublattice 0 in our convention) and the second (belonging to tetrahedral
sublattices 1, 2, and 3) at an angle of about 54.7~degrees to it. Looking at the crystal-field spectra for these sublattices (and for
various sets of crystal-field parameters) one observes that the degeneracy of the ground-state Ising doublet is lifted in arbitrarily
weak fields as one expects - the moments are pinned, as far as possible given the anisotropy, along the field direction.
For sublattice 0, there is a succession of excited-state level crossings, with the ground state being pinned along the field direction.
For the other sublattices, there is a ground-state level crossing (at about 62~T for the parameters of Ref.~\onlinecite{Rosenkranz}) and
a crossing of the first-excited level at about half of this field. The ground-state level crossing is associated with a step in the
magnetization which is not present in the experimental data up to 60~T although there is an upturn in the magnetization at the highest
fields which is also present in the calculated magnetization up to 60~T.
We expect that an exploration of the magnetization at higher field would reveal the predicted step in the magnetization within a few
Tesla of 60~T.

\begin{figure}
\begin{center}
\includegraphics[width=0.77\columnwidth]{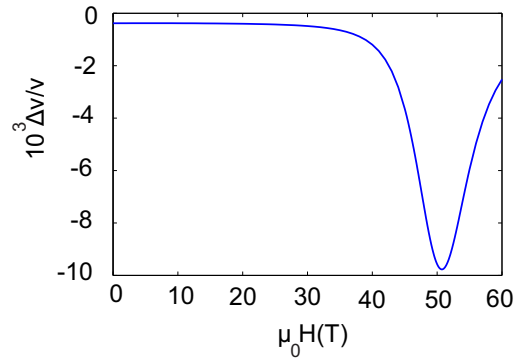}
\end{center}
\caption{Calculated magnetic-field dependence of $\Delta v/v$ in HTO from field parameters of Ref.~\onlinecite{Rosenkranz} and $H \| [111]$.
The dip in the sound velocity is a typical feature - only weakly dependent on the choice of magnetoelastic couplings.}
\label{fig:deltav}
\end{figure}

We have calculated $\Delta v /v$ for the parameters obtained from the point-charge approximation. For a large number of choices of
magnetoelastic couplings, the picture presented by these calculations is that a single feature dominates $\Delta v /v$ with a minimum at
fields below the ground state level crossing by up to 10~T (Fig.~\ref{fig:deltav}). It is, therefore, likely that the ground-state level
crossing on magnetic sites belonging to sublattices 1, 2, and 3  is responsible for the form of the high-field sound-velocity data.
\section{conclusion}
The spin-ice materials HTO and DTO show a number of unusual features in their ultrasound properties. Although the overall temperature dependence for both materials is similar at high temperatures, pronounced differences appear at low temperatures and low magnetic fields (close to and in the
spin-ice state). Anomalies in the sound velocity and sound attenuation in DTO and HTO are observed at around 0.6 and 2~K, respectively,  proving the relevance of the magneto-elastic interactions for physics of these spin-ice materials. Additional evidence for the spin-strain couplings comes from peculiarities in the field-dependent acoustic properties in the spin-ice state.

The field-induced non-equilibrium phenomena in the spin-ice state of DTO appear for different demagnetization factors and
sound-propagation directions, but vanish when there is a strong thermal coupling to the bath. This observation demonstrates the
crucial role of the thermal runaway as a balance between the intrinsic and extrinsic non-equilibrium field-driven processes in spin-ice. In HTO, no thermal runaway was found.

At the gas-liquid-type phase transition in DTO a sharp dip in the sound attenuation appears. This striking anomaly
indicates that in presence of sound waves there may be a dynamical steady state having higher energy resulting in negative relaxation
processes. The experimental results agree qualitatively with our theoretical modeling of these processes, based on
exchange-striction couplings. Remarkably, no such relaxation processes have been observed in HTO.

Pulsed-field ultrasound data exhibit anomalies around 50~T due to CEF effects in both spin-ice materials. Our high-magnetic-field
results are described reasonably well by a CEF model. Calculations based on the CEF scheme have revealed that these anomalies are
related to various level crossings in these spin-ice systems.
Our investigations provide new insight into the role of the lattice degrees of freedom and spin-strain interactions, as well as the
thermal equilibration in spin-ice materials.

\begin{acknowledgments}
We acknowledge the support of the HLD at HZDR, member of the European Magnetic Field Laboratory (EMFL).
A.A.Z. acknowledges the support from the Institute for Chemistry of V.~N.~Karazin Kharkov National University.
Work at Warwick was supported through the EPSRC grant EP/I007210/1.
\end{acknowledgments}

\end{document}